\documentclass[acmsmall,screen,noacm]{acmart}

\renewcommand\footnotetextcopyrightpermission[1]{}

\AtBeginDocument{%
  }

\setcopyright{acmlicensed}
\copyrightyear{2018}
\acmYear{2018}
\acmDOI{XXXXXXX.XXXXXXX}

\acmConference[Conference acronym 'XX]{Make sure to enter the correct
  conference title from your rights confirmation emai}{June 03--05,
  2018}{Woodstock, NY}

\acmISBN{978-1-4503-XXXX-X/18/06}

\usepackage{xspace}
\usepackage{pifont}
\usepackage{tikz}
\usepackage{xcolor}
\usepackage{color}
\usepackage{listings, multicol}
\usepackage{tabularray}
\usepackage[ruled,norelsize,linesnumbered]{algorithm2e}
\usepackage{algpseudocode}
\usepackage{setspace}
\usepackage{hyperref}
\usepackage{tablefootnote}
\usepackage{threeparttable}
 \usepackage{enumitem}
\usepackage[normalem]{ulem}
\usepackage{multirow}

\usepackage{tikz}

\newcommand{\circnum}[1]{%
  \tikz[baseline=(char.base)]{
    \node[shape=circle,draw,inner sep=1pt] (char) {#1};
  }%
}

\begin{document}
\title{Rover: Context-aware Conflict Resolution with LLM}

\author{Qingyu Zhang}
\affiliation{%
  \institution{The University of Hong Kong}
  \country{China}
}
\email{z1anqy@connect.hku.hk}

\author{Junzhe Li}
\affiliation{%
  \institution{The University of Hong Kong}
  \country{China}
}
\email{jzzzli@connect.hku.hk}

\author{Jiayi Lin}
\affiliation{%
  \institution{The University of Hong Kong}
  \country{China}
}
\email{linjy01@connect.hku.hk}

\author{Changhua Luo}
\affiliation{%
  \institution{Wuhan University}
  \country{China}
}
\email{chluo2502@whu.edu.cn}

\author{Chenxiong Qian}
\authornote{Corresponding author.}
\affiliation{%
  \institution{The University of Hong Kong}
  \country{China}
}
\email{cqian@cs.hku.hk}

\newcommand{\sys}{\textsc{Rover}\xspace}
\newcommand\QY[1]{\textbf{\textcolor{blue}{QY: #1}}}
\newcommand\LCH[1]{\textbf{\textcolor{orange}{LCH: #1}}}
\newcommand{\JZ}[1]{{\textbf{\textcolor{magenta}{JZ: #1}}}}
\newcommand\JY[1]{\textbf{\textcolor{orange}{LCH: #1}}}
\newcommand*\blackcircled[1]{\tikz[baseline=(char.base)]{
    \node[shape=circle,draw,inner sep=1pt, minimum size=0.8em] (char) {\textcolor{black}{\scriptsize #1}};
}}
\newcommand{\eg}{{\em e.g.}\xspace}
\newcommand{\etal}{{\em et al.}\xspace}
\newcommand{\etc}{{\em etc.}\xspace}
\newcommand{\ie}{{\em i.e.}\xspace}

\newenvironment{mybullet}{\begin{list}{$\bullet$}
		{\setlength{\topsep}{0.5mm}\setlength{\itemsep}{0.5mm}
			\setlength{\parsep}{0.5mm}
			\setlength{\itemindent}{0.5mm}\setlength{\partopsep}{0.5mm}
			\setlength{\labelwidth}{15mm}
			\setlength{\leftmargin}{4mm}}}{\end{list}}

\makeatletter 
\newcommand{\removelatexerror}{\let\@latex@error\@gobble}
\makeatother

\renewcommand{\algorithmicrequire}{\textbf{Input: }}  % Use Input in the format of Algorithm  
\renewcommand{\algorithmicensure}{\textbf{Output: }}

\lstset{basicstyle=\ttfamily\scriptsize\bfseries, firstnumber=1, xleftmargin=1em, xrightmargin=-0.3cm, numbers=left, numberstyle=\tiny, keywordstyle=\color{blue!70},
commentstyle=\color{red!50!green!50!blue!50}, frame=shadowbox,
rulesepcolor=\color{red!20!green!20!blue!20} }

\begin{abstract}
Code merging is a significant challenge, particularly in large-scale projects. 
Existing solutions, including program analysis and machine learning, show promise but face critical limitations. 
Program analysis lacks the ability to infer developers’ intentions, relying on conservative strategies that offload unresolved conflicts for manual handling. 
Meanwhile, model-based approaches struggle with conflicts involving complex code dependencies due to insufficient contextual awareness. 
To address these gaps, we introduce \sys, a novel conflict resolution system that integrates program analysis with large language models (LLMs). 
To obtain context-aware prompts, we propose Multi-layer Code Property Graph (MtCPG), a new representation capturing inter-file dependencies and enabling contextual analysis for a given conflict.
Using graph connectivity algorithms, \sys further clusters conflicting code and associated changes into meaningful "contexts" that guide the LLM in generating accurate resolutions.
We compared \sys with standalone LLMs, machine learning baseline MergeGen, and suggestion provider tool WizardMerge with adjacent code as the contexts.
Evaluation results show that \sys surpasses all of these approaches in terms of conflict resolution, achieving higher similarity to ground-truth resolutions at character, lexical, and semantic levels.
\end{abstract}




\maketitle

\section{Introduction}

Modern software development heavily relies on version control systems (VCS)~\cite{vcs1, vcs2}, such as Git~\cite{git_vcs} and SVN~\cite{svn_vcs}, to support collaborative development.
As projects evolve in parallel, merging independently modified branches becomes inevitable.
When concurrent changes are incompatible, merge conflicts~\cite{conflict} arise and require manual resolution, which is widely recognized as a time-consuming and error-prone process.
In a previous study, the authors collected datasets from 100 popular repositories on GitHub to assess the challenges of conflict resolution.
The study shows that 75\% of the conflicting merge scenarios take up to 1.77 hours to resolve, while the maximum resolution time consumption is around 530 days~\cite{vale2021challenges}.

Initially, textual and structural merging approaches have been introduced to assist in conflict resolution.
Textual merging~\cite{textual-merge1, textual-merge2} treats merge candidate files as plain strings and relies on string comparison algorithms.
These merging tools are efficient and are applicable to any type of textual file.
However, due to the lack of code-aware analysis, textual merging often compromises correctness and may produce incorrect merge results~\cite{vdcb_1, incorrect_build}.
In contrast, structural merging~\cite{struct-merge1, struct-merge2, spork, mastery, safemerge, automerge} performs static analysis before merging and generally achieves more reliable results.
To combine the advantages of both approaches, semi-structural merging~\cite{semi-structural-merge, jdime, fstmerge, intellimerge} heuristically applies structural merging to code regions and textual merging to non-code content.
Despite these advances, such tools still frequently generate conflicts that require manual resolution.
To further automate conflict resolution, prior studies have explored machine learning–based techniques~\cite{gmerge, MergeBERT, elias2023towards, aldndni2023automatic, deepmerge, mergen}.
After training on curated datasets, these models can resolve certain types of conflicts, but their applicability is often limited to specific languages or conflict patterns.

With the advent of large language models (LLMs)~\cite{llms}, recent work has investigated the use of general-purpose LLMs for conflict resolution~\cite{chatmerge}.
These approaches avoid task-specific pre-training and demonstrate promising results across multiple programming languages.
However, the generation of conflict resolution often requires understanding code dependencies that extend beyond the immediate vicinity of the conflict, including variable definitions, function implementations, and cross-file interactions.
While previous works either did not include the contexts~\cite{chatmerge}, or follow netrual language tasks' idea to treat the adjacent code as code context.
Although neighboring context is effective in natural language tasks, it is often insufficient for code, where the true dependency context may consist of variables, functions, type definitions, or macros defined elsewhere in the same file or even in different files.
Prior research~\cite{congra} shows that simply increasing the length of adjacent context does not consistently improve LLM-based conflict resolution accuracy, highlighting the need for more semantically grounded context selection strategies.
Lacking precise contextual information may lead to unsatisfying resolution quality.
More recently, agent-based approaches have been proposed to address complex software engineering tasks, including conflict resolution, by allowing LLMs to iteratively explore repositories, retrieve relevant files, and refine solutions through multi-step reasoning~\cite{sweagent, repoagent, autocoderover}.
Such agents are, in principle, capable of resolving conflicts by autonomously searching for related code and incorporating broader repository-level context.
Nevertheless, agentic approaches tightly couple context discovery, reasoning, and generation within a non-deterministic exploration loop, making it difficult to guarantee the relevance and completeness of the retrieved context, or to attribute failures to missing context versus reasoning errors.
Moreover, agents incur substantial computational and engineering overhead, and their exploration paths are often unstable and hard to reproduce, which limits their suitability for systematic evaluation.

In this paper, we present \sys, a novel automated conflict resolution system that combines the generative capabilities of LLMs with program analysis.
The design of \sys is guided by two key insights.
First, although LLMs can consume adjacent code blocks as contextual input, they often fail to capture the broader semantic dependencies of a code fragment, such as variables, functions, or composite types defined elsewhere in the project.
We argue that \textbf{these dependencies are essential for correctly understanding the semantics of a conflict, yet they remain largely invisible to LLMs when only limited, localized context is provided}.
Second, while agent-based approaches are capable of retrieving additional context for a given conflict, their known limitations suggest that \textbf{effective conflict resolution should be grounded in a principled and analyzable notion of context rather than heuristic exploration}.
Together, these insights motivate the integration of dependency analysis into LLM-based conflict resolution systems.
However, existing dependency analysis techniques are not well-suited to code merging scenarios.
On the one hand, language-specific tools~\cite{lattner2004llvm, vallee2010soot, bruneton2002asm} lack generality across programming languages and typically require compiling the entire project, even though only a small fraction of files are involved in a merge.
Moreover, when code changes occur in files that are not compiled, such tools may fail to capture the relevant dependencies, leading to incomplete analysis.
On the other hand, language-agnostic approaches~\cite{cpg,tree_sitter} provide only limited dependency information.
For example, \textit{tree\_sitter}~\cite{tree_sitter} operates primarily at the syntactic level and cannot model richer dependencies such as control flow or data flow.
In addition, both categories are designed to analyze a single program version and therefore fall short in capturing dependencies across multiple versions, which is a fundamental requirement in merge conflict resolution.

We choose to implement language-agnostic tools in \sys to cooperate with LLM's multilingual ability.
To address the limitation about lacking of comprehensive code dependency analysis, we introduce a novel architectural component at the core of \sys: the Multi-layer Code Property Graph (MtCPG). 
Atop CPG structure~\cite{cpg}, MtCPG represents the dependency among various definitions and attaches control flow and program dependency relationships to the corresponding items, which allows a comprehensive analysis process. 
Compared with the original CPG, MtCPG can additionally retrieve more dependency on declaration and definition, and support inter-file dependency analysis.
To address the limitation about fitting in merging scenarios, \sys then aligns the nodes of the MtCPG with the conflict and other code change snippets to match the MtCPG with the text information of merging so as to ensure that only the context of the conflict will be analyzed.
After that, \sys employs graph connectivity algorithms to group conflict blocks and associated code changes into distinct clusters. 
These clusters ensure that all relevant code dependencies are accurately identified and grouped together, providing the LLM with a rich, semantically meaningful context for conflict resolution.
Finally, \sys synthesizes the code dependence context along with the conflicting code and utilizes this comprehensive understanding to guide the LLM in generating precise and automated conflict resolutions.

To evaluate \sys, we conducted experiments to assess how MtCPG outperforms CPG and \sys's effectiveness in improving conflict resolution accuracy.
The results demonstrate that MtCPG can increase the number of the graph's nodes and edges by 3.4\% and 25\% compared with the original CPG.
Also, \sys outperforms a standalone LLM, \textit{Qwen3-30B-A3B} by 13.29\%, 10.46\%, and 9.79\% in terms of lexical, character, and semantic similarities on ConGra's dataset~\cite{congra}. 
Compared to MergeGen ~\cite{mergen}, \sys achieves improvements of 17.72\%, 15.90\%, and 9.10\% respectively. 
Additionally, \sys surpasses the combination of LLM with WizardMerge~\cite{wmerge}, achieving better performance
by 3.82\%, 1.85\%, and 4.07\% respectively on WizardMerge's dataset.
These results indicate that proper contextual information is potential to improve LLM's conflict resolution generation.
%
%They suggest that explicitly modeling and leveraging code dependencies is a key factor in enabling LLMs to produce conflict resolutions that better preserve the intent of developers’ changes, thereby reducing the manual effort required during real-world code merging.

%
In conclusion, we make the following contributions:

\begin{mybullet}
    \item We proposed MtCPG, a novel representation scheme to represent the dependence for all code components.
    \item We proposed a remarkable methodology to classify each code block with changes based on the MtCPG.
    On top of this, code contexts, which are way more suitable for assessing the reliance of code compared with adjacent code will be generated. 
    \item We implemented \sys, an LLM-based conflict resolving tool atop the above approaches, and evaluated it on ConGra along with a standalone LLM to prove its effectiveness.
    \sys will be public at \url{https://figshare.com/s/fa83d33b5759da38a1a1}.
\end{mybullet}
\section{Background}

\subsection{Code Merging Automation}
In a prior study conducted by Vale \etal, the authors show that the complexity of merge conflicts—measured by factors such as the number of involved files and lines of code—has a significant impact on the time developers spend resolving them~\cite{vale2021challenges}.
It highlights that conflict resolution is not merely a local editing task, but often requires understanding code that spans multiple locations and abstraction levels.
For projects with extremely large codebases, such as the Linux kernel, manually resolving merge conflicts therefore poses a substantial challenge to developers.
Such manual resolution processes are typically time-consuming, error-prone, and inefficient.

Textual-level merging approaches were first implemented in version control systems~\cite{textual-merge1, textual-merge2}.
These methods treat each file as plain text, offering strong generality and high efficiency.
However, purely textual merging often produces incorrect results when semantic dependencies are involved.
To mitigate these issues, researchers have proposed structural-level merging approaches that incorporate static analysis~\cite{struct-merge1, struct-merge2, spork, mastery, safemerge, automerge}.
Such tools typically translate source code into structured representations (\eg, abstract syntax trees) and perform merging at the graph level.
While structural merging can detect syntactic and certain semantic conflicts more accurately, it primarily focuses on conflict detection rather than automated resolution or systematic identification of the affected semantic context.
To balance effectiveness and efficiency, semi-structured merging methods have also been proposed~\cite{semi-structural-merge, jdime, fstmerge, intellimerge}.
These approaches apply structural merging to code regions and textual merging to non-code content.
Nevertheless, unresolved conflicts that require manual intervention remain common.
To further automate conflict resolution, prior work has explored machine learning–based approaches~\cite{gmerge, MergeBERT, elias2023towards, aldndni2023automatic, deepmerge, mergen}.
By training models on curated conflict-resolution examples, these methods can generate resolutions for specific conflict patterns, thereby assisting developers during merging.
However, such approaches typically require task-specific pre-training and large amounts of labeled data.
The emergence of LLMs provides an alternative by enabling conflict resolution without task-specific training~\cite{chatmerge}.
In these approaches, conflict code blocks are embedded into prompts, and LLMs generate corresponding resolutions.
Despite their flexibility, existing LLM-based methods largely rely on adjacent code blocks as context, which often fails to capture the true semantic dependencies of the conflict.
More recently, agent-based systems~\cite{sweagent,repoagent,autocoderover} have been proposed to enrich context by autonomously exploring repositories.
These agents iteratively infer relevant files and code regions and expand context based on the current repository state.
While promising, such approaches introduce new challenges in terms of determinism, interpretability, and evaluation, which are further discussed in the following section.

\subsection{Limitation of Existing Works}\label{sec:lims}

\begin{figure}[th]
	\centerline{\includegraphics[width=0.8\linewidth]{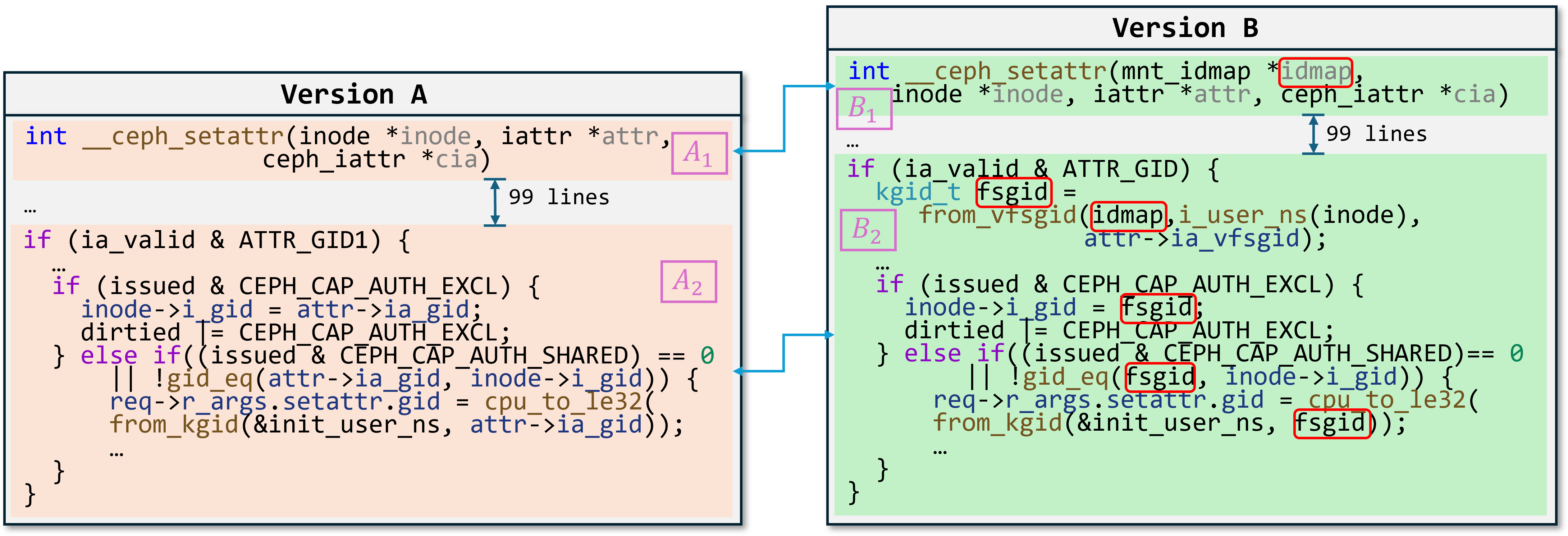}}
	\caption{\textmd{A code context reliance example.}}
	\label{fg:far_ctx}
\end{figure}

Although extensive research has been conducted on code merging and conflict resolution, existing approaches still face fundamental limitations that hinder fully automated and reliable resolution.
On the one hand, neither textual nor structural merging tools can eliminate all conflicts, leaving developers responsible for manually resolving unresolved cases.
On the other hand, machine learning–based approaches require task-specific pre-training and large amounts of conflict data, which limits their scalability and practical applicability.

For LLM-based approaches, a major limitation lies in context selection.
Most existing methods rely on adjacent code blocks as context, which may be inefficient or even misleading.
According to the ConGra benchmark~\cite{congra}, increasing the length of adjacent context does not necessarily improve conflict resolution accuracy, indicating that context relevance is more critical than context size.

We provided an example of code reliance in ~\autoref{fg:far_ctx}.
In file \textit{fs/ceph/inode.c} of linux kernel project, \textit{Version A} and \textit{Version B} are different in two pairs of code blocks (\ie $A_1-B_1$ and $A_2-B_2$).
For the first pair of changes, \textit{Version B} adds a new parameter \textit{idmap} for the function prototype \textit{\_\_ceph\_setattr}.
In the second pair of changes, the codes from \textit{Version A} and \textit{Version B} differ because the new parameter \textit{idmap} is used to update (\textit{fsgid} and \textit{inode->i\_gid}) and is passed to functions \textit{gid\_eq} and \textit{from\_kgid}.
Since the parameter \textit{idmap} used in the second pair is defined in the first pair, any conflict in the first pair of changes necessitates a re-evaluation of the second pair, regardless of whether it has conflicts. 
In other words, the two pairs of code changes are related, even though they are separated by 99 lines in the code. 
However, in existing research about conflict resolution LLMs, the context always refers to the adjacent code blocks around the conflict.
While a small context line fails to cover such a distance between code blocks, a large context line may exceed the limitation of the token numbers predefined by the models.
Moreover, if the dependence code block is in another file (\eg creating a variable with a type that is defined in another file), employing a larger context size will not help with increasing the accuracy as it cannot achieve inter-file context awareness.

Although agent-based tools can reach a broader context, they face inherent challenges when applied to conflict resolution.
Their exploration process is driven by LLM decisions, resulting in non-deterministic and potentially incomplete context selection.
Furthermore, context discovery, reasoning, and generation are tightly coupled, making it difficult to attribute resolution failures to missing context or flawed reasoning.
The associated computational overhead also limits scalability.

Motivated by these limitations, we argue that conflict resolution requires a principled and analyzable notion of context.
Instead of relying on LLM-driven exploration, our approach leverages program analysis to deterministically identify conflict-centric semantic context, and then conditions LLM-based resolution on this context.
This design choice forms the foundation of \sys.
Plus, our work provides a deterministic and analyzable foundation for context selection, which can potentially benefit future agent-based systems.
\section{Design}

To address the above limitations, the design of \sys needs to consider the following two challenges:

% \begin{mybullet}
\textbf{Challenge 1: Inefficiency in Code Dependency Analysis.}
Existing approaches to dependency analysis can be categorized into language-specific and language-agnostic methods.
Language-specific techniques~\cite{lattner2004llvm, vallee2010soot, bruneton2002asm} transform source code into an intermediate representation for analysis.
However, these methods lack generalizability across different programming languages and require the entire project to be compiled beforehand, despite the fact that files with code changes often affect only a small portion of the project.
Furthermore, codes that are not compiled remain unanalyzed, potentially leading to missing dependencies.
If code changes reside in uncompiled code segments, the resulting dependency analysis will be incomplete.
Language-agnostic approaches~\cite{cpg, tree_sitter} fail to provide a comprehensive analysis of all types of code dependencies.
For instance, tree\_sitter constructs syntax trees from code but struggles with identifying data dependencies.
Consequently, if a conflict depends on other code blocks via data dependencies, tree\_sitter fails to capture this contextual information, leading to an incomplete dependency representation.

\textbf{Challenge 2: Misalignment Between Text Representations and Dependency Structures.}
Given that code changes between two merging versions are represented in text format, the first step is to align these textual modifications with their corresponding elements in the dependency representation.
A straightforward mapping algorithm suffers from high computational complexity, making it inefficient for large-scale projects.
Furthermore, once dependency analysis is performed with code change information incorporated, the results must be translated back into textual form to serve as contextual information for prompting LLMs.
However, an effective strategy for synthesizing the dependencies of both versions into a unified contextual representation remains unknown.
% \end{mybullet}

\subsection{Overview}\label{sec:overview}

\begin{figure*}[th]
	\centerline{\includegraphics[width=\linewidth]{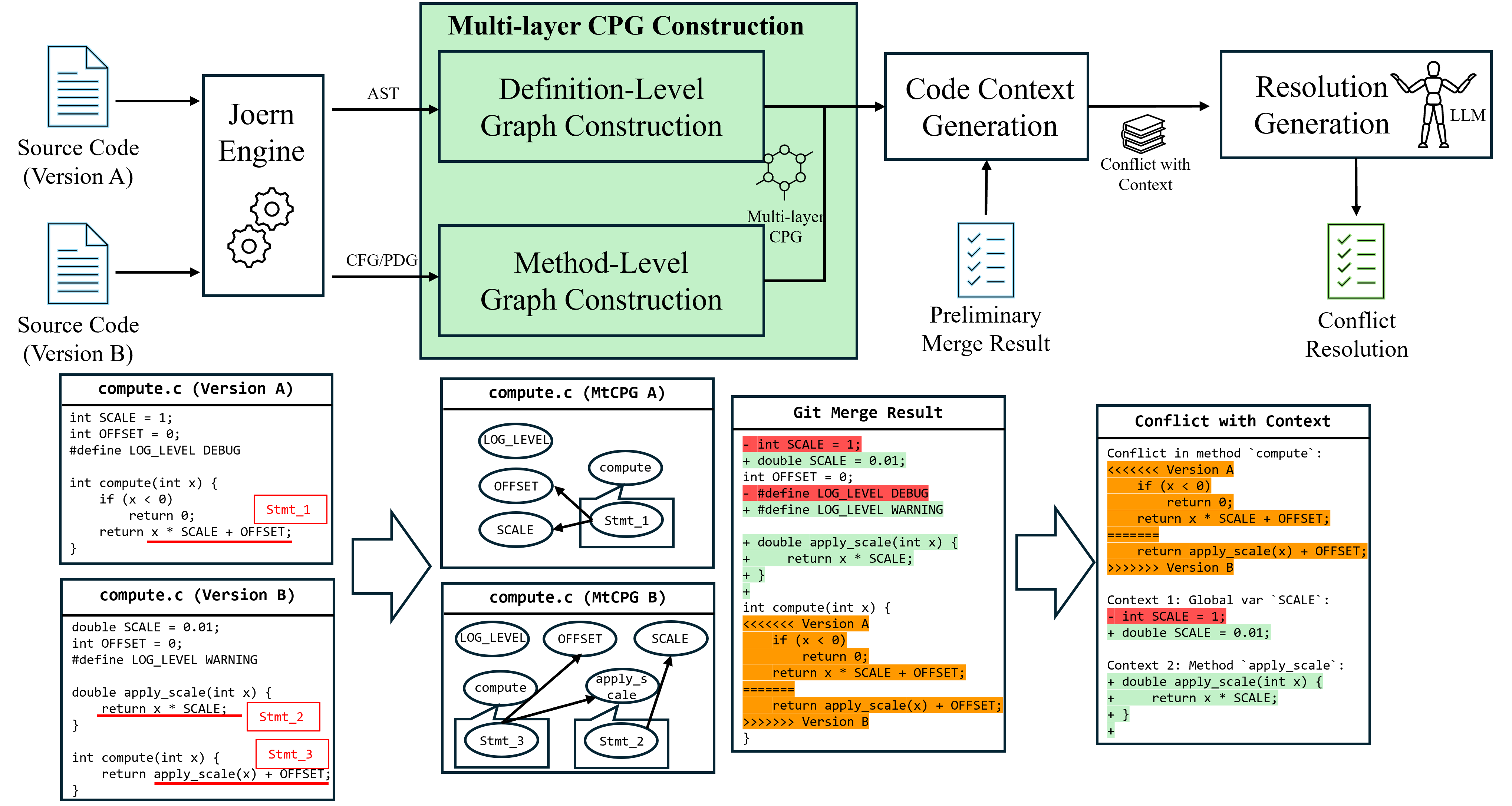}}
	\caption{\textmd{Overview of \sys System.}}
	\label{fg:overview}
\end{figure*}

We present the overview of \sys in ~\autoref{fg:overview} together with a concrete code example to explain how \sys addresses the aforementioned challenges.
First, \sys consumes the files that contain changes and feeds them into a source code analyzer to achieve language diversity and generate an initial CPG. 
Using the initial CPG as the representation, \sys can analyze code files written in any language.
We choose Joern~\cite{cpg} as it has implemented a mature framework for CPG generation.
Then, \sys targets the first challenge by constructing the Multi-layer Code Property Graph (MtCPG) to represent both the dependence between definitions and the control flow, data flow, and program dependence derived from CPG.
After the former step, \sys can obtain two MtCPGs for both of the merging candidate versions.
To address the second challenge, we implemented a coloring mechanism to attach each node on MtCPG with the corresponding source code information from preliminary merge results.
\sys then analyzes the dependence within the MtCPG by its connectivity to generate contexts for each conflict.
Ultimately, the conflict, along with the context codes, will be fed into the LLM for automatic resolution generation with our pre-defined prompt.

\subsection{Multi-layer Code Property Graph}\label{sec:mcpg}

%
% There are two alternative ways to achieve the code dependency analysis, \ie language-specific and language-agnostic. 
% %
% Language-specific approach requires compiling the whole project to dump the IR and perform analysis. 
% %
% It is more time-consuming and less general among languages.
% %
% Moreover, It will miss the dependency from codes that are not compiled.
% %
% On the other hand, language-agnostic approach executes faster and can be easily migrated to other languages (\eg CPG-based tools ~\cite{cpg1, cpg2, cpg3}). 
% %
% Nevertheless, their analysis accuracy will be compromised. 
% %
% For example, when a variable is defined with a type, a reliance edge should be built from the definition of the variable to the definition of the type.
% %
% However, the node of the variable's definition in the current CPG only records the name of the dependent type in a string, but no edge directly linked to the definition node of the type itself.
% \LCH{I think we should elaborate more on the unique challenges of using Joern to solve conflicts (e.g., we do not analyze a regular program but a program with conflict code changes), the above reads like general limitations. And not clear why this (and also the below paragraph) is important for solving conflicts}

%
We propose \textbf{MtCPG} to address \textbf{Challenge~1} by enabling both generality and accuracy in \sys.
Our approach starts from a vanilla Code Property Graph (CPG) constructed by Joern, which provides a reliable foundation for capturing intra-procedural dependencies at the source-code level.
Since \sys only requires basic CPG construction without additional semantic analyses, this step introduces minimal overhead and remains efficient in practice.
Building upon the vanilla CPG, MtCPG introduces an expressive extension that fundamentally differs from traditional CPGs in its dependency modeling.
Specifically, MtCPG explicitly incorporates \emph{definition-level abstractions}, \emph{cross-layer dependencies}, and \emph{inter-file dependencies}, which are either implicit or entirely absent in Joern’s original CPG.
By organizing program elements into hierarchical layers and connecting them via cross-layer edges, MtCPG enables fine-grained dependency reasoning between entire definitions and their internal components.
Moreover, by introducing inter-file edges mediated through import operations, MtCPG explicitly captures cross-file dependency relations without flattening or merging source files.
These extensions allow MtCPG to support precise and scalable dependency reasoning in real-world, multi-file projects, while preserving the generality and efficiency of the original CPG.

\begin{table}
\centering
\caption{\textmd{Node types in MtCPG, including their definitions and representative code examples.}}
\label{tb:mtcpg_nodes}
\resizebox{\linewidth}{!}{\tiny
\begin{tabular}{c|c|c}
\hline
\textbf{Node Name} & \textbf{Definition} & \textbf{Code Example} \\ \hline
TypeDef
& Composite type definitions and type aliases
& \texttt{struct c\_type\_s \{...\};} \\

MethodDef
& Method/function definitions and function-like macro definitions
& \texttt{int foo(...)\{...\}} \\

MemberDef
& Struct or class member definitions
& \texttt{struct \{ int a; \};} \\

GlobalVarDef
& Global variable definitions and object-like macro definitions
& \texttt{\#define PI 3.14} \\

ImportDef
& External component inclusion
& \texttt{\#include "bar.h"} \\

MethodStmt
& Statements inside a method body
& \texttt{int ...(...)\{a = b + c;\}} \\

MethodVarDef
& Parameters and local variables in a method
& \texttt{int ...(...)\{int a;}\} \\

\hline
\end{tabular}}
\end{table}

\sys's current design incorporates seven types of MtCPG nodes, as summarized in~\autoref{tb:mtcpg_nodes}.
These nodes represent different kinds of program definitions and structural elements.
For example, \textit{GlobalVarDef} represents global variable definitions as well as \emph{object-like macro} definitions, while \textit{MethodDef} represents function and method definitions together with \emph{function-like macro} definitions.
Note that although macros are expanded during preprocessing, we model them as definition nodes to uniformly capture the static dependencies they introduce, which aligns with \sys's focus on dependency analysis rather than execution semantics.
\textit{ImportDef} abstracts language-specific inclusion mechanisms (e.g., \texttt{\#include} in C/C++ or \texttt{import} statements in other languages), serving as a bridge between files and enabling inter-file dependency analysis.
\textit{MethodStmt} denotes individual statements within a method body, allowing MtCPG to capture fine-grained intra-procedural dependencies, while \textit{MethodVarDef} represents local variables and function parameters.

\begin{figure}[th]
	\centerline{\includegraphics[width=0.6\linewidth]{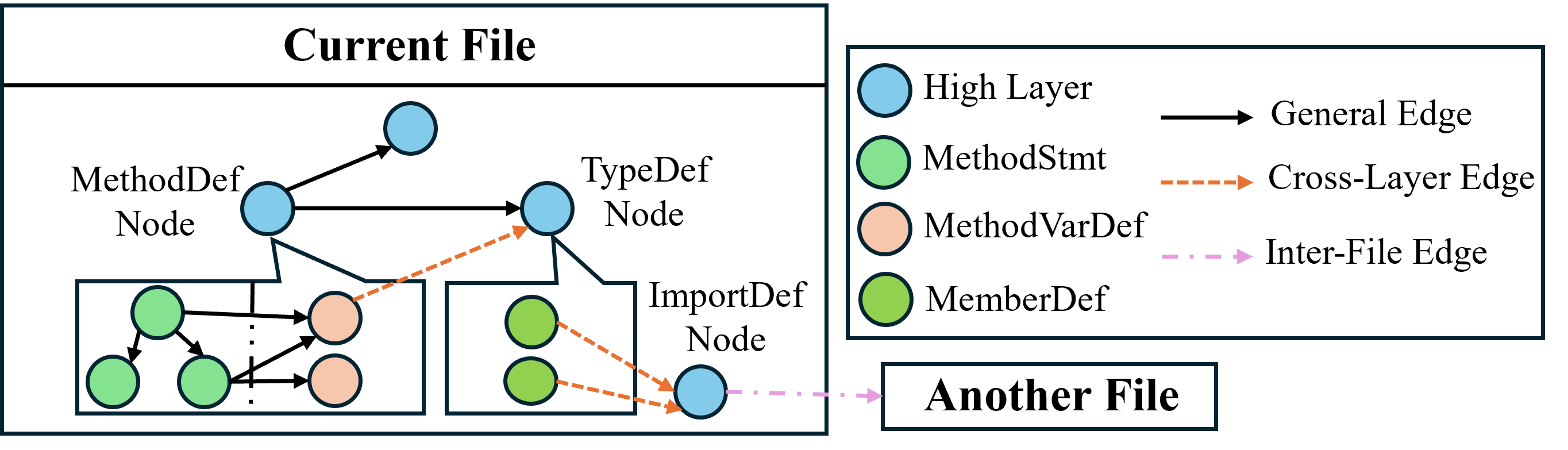}}
	\caption{\textmd{Concept graph of multi-layer code property graph.}}
	\label{fg:concept}
\end{figure}

To better illustrate the high-level design of MtCPG and highlight its differences from traditional CPGs, we use~\autoref{fg:concept} to explain how MtCPG is constructed.
Unlike conventional CPGs, which mainly focus on intra-procedural or single-file dependencies, MtCPG explicitly introduces \emph{cross-layer edges} and \emph{inter-file edges} to enable both finer-grained and cross-file dependency reasoning.
In MtCPG, \sys organizes nodes according to the structural hierarchy of source code, forming two conceptual layers.
Specifically, \textit{TypeDef}, \textit{MethodDef}, \textit{GlobalVarDef}, and \textit{ImportDef} are classified as \emph{high-layer nodes}, as they represent standalone definitions or file-level program elements.
In contrast, \textit{MemberDef}, \textit{MethodStmt}, and \textit{MethodVarDef} are treated as \emph{low-layer nodes}, since they are structurally contained within corresponding high-layer definitions (e.g., a \textit{MethodStmt} node belongs to a specific \textit{MethodDef} node).
This hierarchical organization allows \sys to distinguish whether a merge conflict affects an entire definition or only a specific part of it.
Dependencies among high-layer nodes capture definition-level relationships, such as type usage or method invocation across different components.
Meanwhile, low-layer nodes are attached to their corresponding \textit{MethodDef} nodes, and the connections among \textit{MethodStmt} nodes preserve the original CPG dependencies derived from Joern.
These include AST, CFG, or PDG dependencies from the original CPG, together with precise data-flow information (e.g., accessed variables).
Besides the dependence edge from CPG, we additionally introduce the following two types of edges:

\textbf{Cross-layer edges.}
A \textit{cross-layer edge} is constructed whenever a dependency exists between a high-layer node and a low-layer node.
By explicitly linking definition-level abstractions with statement-level elements, cross-layer edges enable MtCPG to perform finer-grained and more exact dependency analysis than traditional CPGs, which typically lack such hierarchical connections.

\textbf{Inter-file edges.}
An \textit{inter-file edge} is constructed when a dependency between nodes residing in different source files is detected via an \textit{ImportDef} node.
For example, if \sys cannot locate the required \textit{TypeDef} node for a \textit{MemberDef} node within the current file, it searches the \textit{TypeDef} nodes in files imported by the current file.
Once a matching \textit{TypeDef} node is identified, \sys first constructs a cross-layer edge from the dependent node to the corresponding \textit{ImportDef} node, indicating that the dependency is introduced through an import operation.
Subsequently, an inter-file edge is added from the \textit{ImportDef} node to the target \textit{TypeDef} node, explicitly representing the cross-file dependency.
Through this two-step construction, MtCPG not only enables inter-file dependency analysis but also preserves the rationale behind each import, i.e., which specific definitions are required from the imported files.
Although a source file may import many other files, potentially leading to high analysis overhead, \sys restricts inter-file exploration to changed files only.
This design significantly reduces the search space while preserving the correctness of dependency reasoning required for accurate merging.

\begin{figure}[th]
	\centerline{\includegraphics[width=\linewidth]{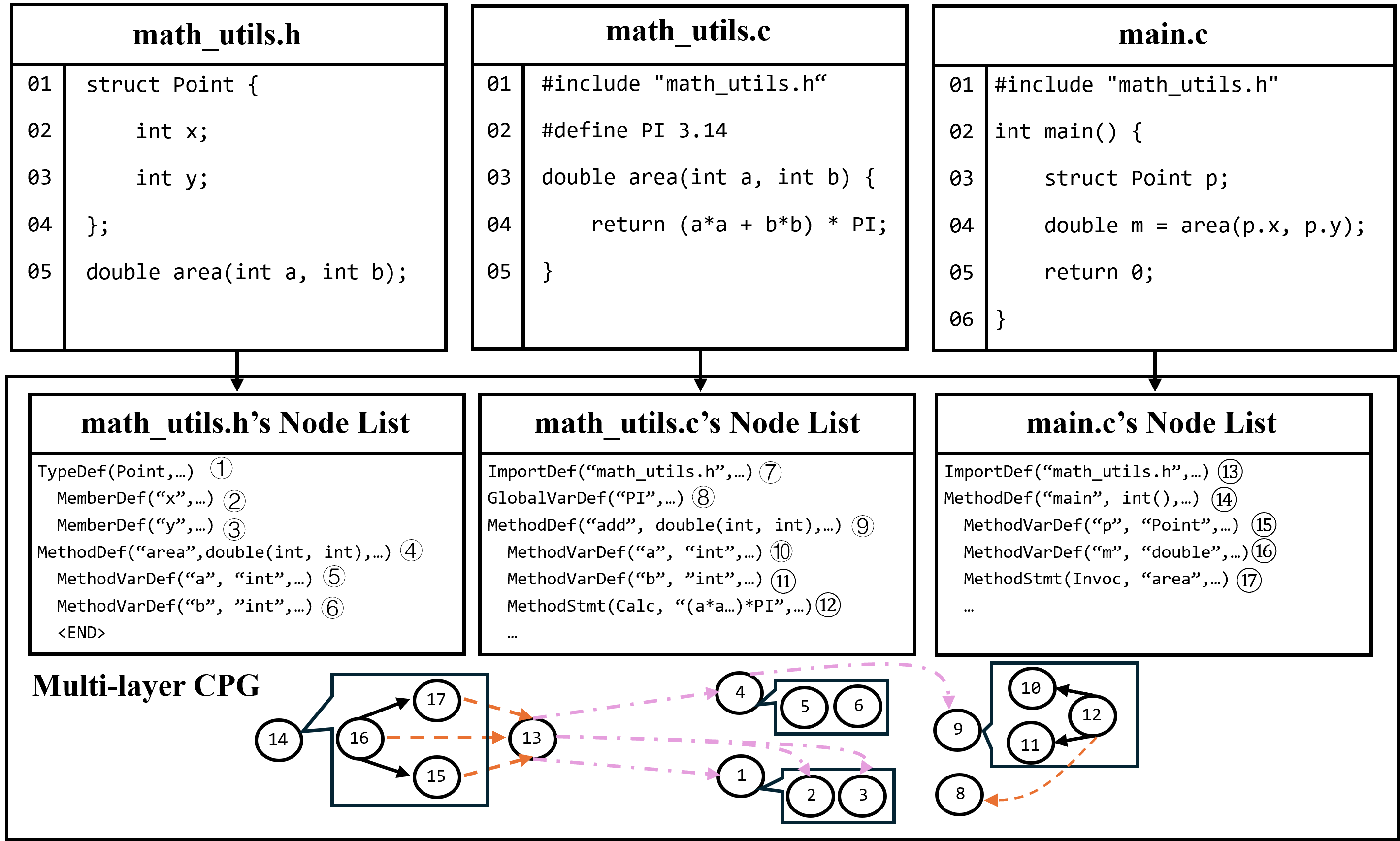}}
	\caption{\textmd{Construction example of multi-layer code property graph.}}
	\label{fg:mtcpgexample}
\end{figure}

We use~\autoref{fg:mtcpgexample} as a running example to illustrate how a set of source files is transformed into MtCPG.
The program first defines a \texttt{Point} structure in \texttt{math\_utils.h} to represent the coordinates of a point.
It then implements a function for computing the area of a circle whose radius is the distance from the point to the origin in \texttt{math\_utils.c} (i.e., the \texttt{area} method), and finally invokes this method in \texttt{main.c}.
During MtCPG construction, \sys parses each source file and generates the corresponding node lists, and constructs CPG edges based on the original CPG produced by the Joern engine (denoted by black solid arrows).
For instance, line~4 in \texttt{math\_utils.c} depends on the two parameters defined at line~3 (i.e., \circnum{12} depends on \circnum{10} and \circnum{11}), which is directly derived from Joern’s original CPG.
Afterward, \sys retrieves cross-layer dependencies and constructs additional edges according to the identified dependency relations (denoted by orange dashed arrows).
For example, the constant \texttt{PI} defined at line~2 of \texttt{math\_utils.c} is used by the statement at line~4.
Accordingly, \sys constructs a cross-layer edge from \circnum{12} to \circnum{8}.
Finally, since \texttt{area} and \texttt{Point} are not defined in \texttt{main.c}, resolving these dependencies triggers inter-file analysis.
In this process, a cross-layer edge is first constructed from the dependent node to the corresponding importing node in the dependent file, followed by an inter-file edge (denoted by purple dashed arrows) from the importing node to the actual dependency node.
Specifically, line~3 in \texttt{main.c} defines a variable of type \texttt{Point} (i.e., \circnum{15}).
This structure type is defined in \texttt{math\_utils.h}, which corresponds to the import node denoted as \circnum{13}.
Thus, a cross-layer edge is constructed from \circnum{15} to \circnum{13} to indicate that the \texttt{Point} type is introduced via the import operation.
In addition, an inter-file edge is constructed from \circnum{13} to \circnum{1}, indicating that the import operation enables access to the \texttt{Point} structure.
Note that in C/C++ programs, a function prototype may be declared in a header file, while its definition resides in a different source file.
To handle this case, we adopt a conservative strategy by constructing inter-file edges from a method declaration to all method definitions whose function signatures match (e.g., from \circnum{4} to \circnum{9}).

\subsection{Code Context Generation}\label{sec:context_generate}

To tackle \textbf{Challenge 2}, \sys takes the next step into extracting the code contexts of each conflict based on the preliminary merge results.
We opt for the merge results from Git because they specify all conflicts and changed code with precise locations.
We face two challenges in this stage.
First, the preliminary merge results reflect the results at the text level, we need to align them with the MtCPG, which is in the format of a graph.
Second, as the MtCPG contains the analysis results of two merging candidate versions, we have to design an accurate and reliable approach to combine them to generate the context. 

\begin{figure*}[th]
	\centerline{\includegraphics[width=\linewidth]{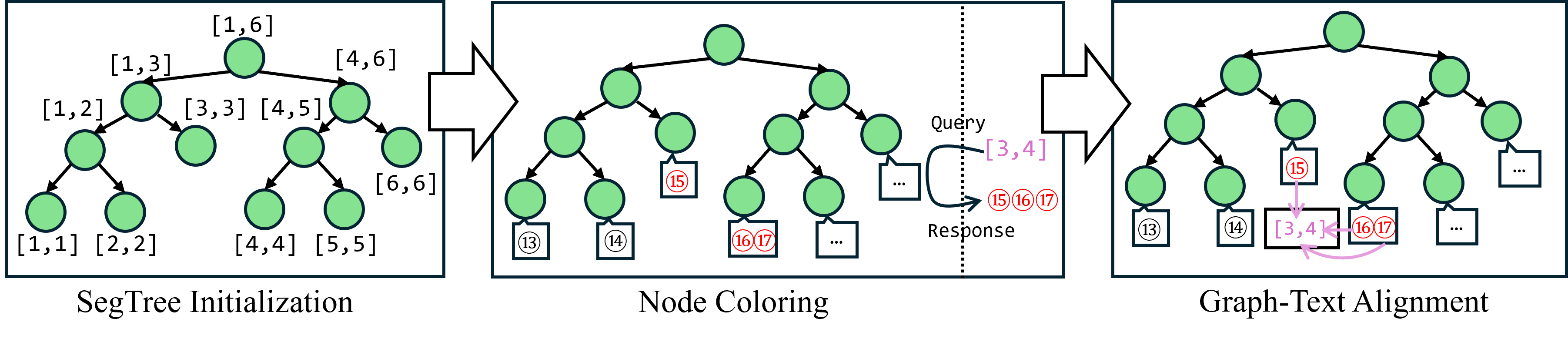}}
	\caption{\textmd{The workflow of \sys's graph-text alignment.}}
	\label{fg:alignment}
\end{figure*}

To align the preliminary merge results with the MtCPG, we extend the "coloring" idea from WizardMerge~\cite{wmerge}. 
In WizardMerge's graph-text alignment, a segment tree~\cite{segmenttree} is used to manage all definition nodes, with modified code blocks serving as the colors to fill these nodes. 
However, this method can fragment one code block into multiple pieces in the preliminary results, while non-colored code is excluded from further analysis. 
In ~\autoref{fg:alignment}, we present \sys's graph-text alignment workflow with a concrete example stemming from ~\autoref{fg:mtcpgexample}'s \texttt{main.c}.
First, we initialize the segment tree for each file with the size of it set to the line number of the file.
For instance, \texttt{main.c} has six lines, so the segment tree is initialized with a range from one to six. 
Then, we annotate each segment tree node with the corresponding MtCPG nodes based on the line-range information associated with each MtCPG node.
Specifically, MtCPG nodes \circnum{13}, \circnum{14}, and \circnum{15} are located at lines 1, 2, and 3, respectively; therefore, the segment tree nodes covering these lines are annotated with the corresponding MtCPG nodes.
In addition, MtCPG nodes \circnum{16} and \circnum{17} are both located at line 4, and consequently, the corresponding segment tree node is annotated twice to reflect both location relationships.
After locating all MtCPG nodes, we construct a query for each code block in the preliminary merge result based on the start and end line numbers of the block.
Given the queried segment tree nodes, \sys returns the MtCPG nodes associated with the corresponding line-range coverage.
For example, in \texttt{main.c}, if lines 3–4 are modified according to the preliminary merge result, a query is issued to the segment tree to retrieve all MtCPG nodes located within this line range (i.e., \circnum{15}, \circnum{16}, and \circnum{17}).
Finally, the diff code block is attached to the corresponding MtCPG nodes sequentially, enabling subsequent context generation to be performed directly over the MtCPG.
In the example file, the three identified MtCPG nodes are all associated with the diff code block \texttt{[3,4]}, indicating that their code dependencies may influence this block.

\begin{figure}[ht]
    \begin{minipage}[t]{0.48\textwidth}
        \removelatexerror
        {\tiny % 调整字体大小
        \begin{algorithm}[H]
            \caption{Single-version context}
            \label{alg:ctx_generate}
                \algrenewcommand{\algorithmicend}{\ }
                \algrenewcommand{\algorithmicthen}{\ }
                \algrenewcommand{\algorithmicdo}{\ }
                \algorithmicrequire Distance Threshold $k$, MtCPG $G$

                \algorithmicensure Classified Code Block Sets $R$       

                \For{$v \in G.nodes$}
                {
                    \uIf{contains\_query\_code\_block($v$)}
                    {
                        $QNodes.add(v)$\;
                    }
                }

                $KQnodes \xleftarrow{} nearest\_k(QNodes, G, k)$\; 

                \For{$v \in QNodes$}
                {
                    \For{$u \in KQnodes[v]$}
                    {
                        \uIf{$v$ is conflict node}
                        {
                            $NodeR.union(v,u)$\;
                        }
                    }
                }

                $T \xleftarrow{} get\_groups\_from\_set(NodeR)$\;

                \For{$t \in T$}
                {
                    \For{$v \in t$}
                    {
                        \For{$cb \in v.code\_blocks$}
                        {
                            \uIf{$r \neq None$}
                            {
                                $r \xleftarrow{} cb$\;
                            }
                            \Else
                            {
                                $R.union(r, cb)$
                            }
                        }
                    }
                }

        \end{algorithm}
        }
    \end{minipage}
    \hfill
    \begin{minipage}[t]{0.48\textwidth}
        \removelatexerror
        {\tiny % 调整字体大小
        \begin{algorithm}[H]
            \caption{Cross-versions context}
            \label{alg:cv_ctx_generate}
                \algrenewcommand{\algorithmicend}{\ }
                \algrenewcommand{\algorithmicthen}{\ }
                \algrenewcommand{\algorithmicdo}{\ }
                \algorithmicrequire Single-version context generation results from two versions $R_A$ and $R_B$, Code block pairs $G$

                \algorithmicensure Cross-versions context generation result $R_{all}$       

                $R_{all} \xleftarrow{} \emptyset$

                \For{$g \in R_A \sqcup R_B$}
                {
                    $r \xleftarrow{} g[0]$\;

                    \For{$cb \in g$}
                    {
                       $R_{all}.union(r, cb)$\; 
                    }
                    
                }

                \While{True}
                {
                    $do\_merge \xleftarrow{} False$\;
                    \For{$cb_A, cb_B \in G$}
                    {
                        $r_A \xleftarrow{} R_{all}.find\_root(cb_A)$\;
                        $r_B \xleftarrow{} R_{all}.find\_root(cb_B)$\;
                        \uIf{$r_A \neq r_B$}
                        {
                            $R_{all}.union(r_A, r_B)$\;
                            $do\_merge \xleftarrow{} True$\;
                        }
                    }
                    \uIf{$do\_merge == False$}
                    {
                        $break$\;
                    }
                }

        \end{algorithm}
        }
    \end{minipage}
\end{figure}

To integrate code context from both versions, \sys first generates context independently for each version based on its MtCPG, as shown in~\autoref{alg:ctx_generate}.
After graph-text alignment, a subset of MtCPG nodes is annotated with query code blocks, indicating their association with code changes.
\sys first collects all such nodes as $QNodes$ (lines 3-6).
To identify related nodes, \sys performs a multi-source breadth-first search (MS-BFS)~\cite{msbfs} starting from all $QNodes$ (line 7), limiting the exploration to at most $k$ steps.
This bound is introduced to (1) exclude distant and less relevant dependencies and (2) avoid excessive context that may degrade LLM performance due to attention sink effects~\cite{gu2024attention}.
The parameter $k$ is user-configurable.
Two $QNodes$ are merged into the same group if they are mutually reachable within $k$ steps and at least one corresponds to a conflict node (lines 8-13).
Nodes without attached conflict code blocks are excluded, as \sys focuses exclusively on conflict-centric context.
Query code blocks are then grouped according to the grouping of their corresponding $QNodes$ (lines 15-25), producing version-specific context groups.
These groups are subsequently combined across versions using the procedure in~\autoref{alg:cv_ctx_generate}.
Specifically, \sys initializes a joint disjoint set $R_{all}$ from the per-version results $R_A$ and $R_B$ (lines 4-9).
It then iterates over each pair of matching code blocks across versions and merges their groups if they are not already connected (lines 10-18).
The process terminates when no further cross-version merges are possible, yielding the final context grouping (lines 19-20).

\subsection{Resolution Generation}\label{sec:res_generate}
At this stage, \sys utilizes both the code context and the conflicting code to guide the LLM in automatically generating a resolution.
The prompt contains six sections in total, which are designed to instruct the LLM to resolve the conflict in alignment with our specific objectives.
The first section is the \textit{Preamble} of the whole task, which introduces the overarching task, specifies the role the LLM should assume, and provides a brief overview of conflict resolution.
Section 2 offers a precise description of the task and clearly defines what the LLM is expected to accomplish (\ie conflict resolution).
The next chain-of-thought section guides the LLM through a structured reasoning process to address the conflict. 
By delineating the thought process in clear, sequential steps, the chain-of-thought methodology enhances the model’s ability to understand intricate problems and solve multi-step reasoning tasks~\cite{cot}.
\sys's chain-of-thought consists of three crucial parts:
1) The LLM must thoroughly analyze the conflict and its surrounding context to identify the root cause of the issue;
2) The LLM needs to infer the intended code logic for both versions, focusing on the high-level differences in functionality introduced by the changes;
3) The LLM must determine the best way to merge the two versions. This could involve adopting one version as is, combining elements from both, or introducing entirely new code logic;
After that, section 4 specifies input and output formats to ensure consistency and clarity.
Instead of providing the entire codebase, we use the "diff" or patch format, as it is more streamlined and easier for the LLM to process and understand~\cite{deligiannis2023fixing}.
To further clarify how to resolve the task, the fifth section provides a concrete example.
It demonstrates how to parse the input in the given format, analyze the conflict using the chain-of-thought process, and generate the resolution while adhering to the required output format. 
This example serves as a reference to guide the LLM's reasoning and output generation.
In the final section, \sys combines the generated code contexts with the conflicting code and feeds them to the LLM as input.

\section{Evaluation}

In this section, we conduct evaluation to answer the following three research questions:
\begin{mybullet}
    \item \textbf{RQ1: }How effectively does \sys’s MtCPG extend the original CPG in capturing dependencies?
    \item \textbf{RQ2: }To what extent does \sys improve the accuracy of LLM-based conflict resolution compared to using standalone LLMs and the baseline tool?
    \item \textbf{RQ3: }How does the performance of \sys compare with alternative tool-assisted LLM pipelines in terms of conflict resolution accuracy?
\end{mybullet}

% \paragraph{Hardware Setting}
% %
% We conducted all the experiments on a server with an Intel(R) Xeon(R) Gold 5418Y CPU (24 cores / 96 threads) and 503GB memory.
% %
% We deployed the selected LLM and baseline model on four 4090 GPUs (24GB).

\textbf{Implementation Detail}
We implemented \sys's program analysis part with an existing analysis framework \textit{Joern 4.0.98}~\footnote{\url{https://github.com/joernio/joern/releases/tag/v4.0.98}}.
We selected \textit{Qwen3-30B-A3B-Instruct-2507}~\cite{yang2025qwen3} as the resolution generation LLM due to its availability on Hugging Face and well-balanced trade-off between performance and cost.
Additionally, to evaluate the generability of \sys across models, we selected \textit{gpt-oss-20b}~\cite{agarwal2025gpt} and \textit{Llama-3.1-8B-Instruct }~\cite{grattafiori2024llama} as the comparison groups.
We set the temperature to 0 to enforce deterministic decoding, which is essential for reliable conflict resolution.

\textbf{Comparison Baseline and Resolution Suggestion Provider Selection}
We chose MergeGen~\cite{mergen} as the machine learning approach baseline to compare with \sys and other control groups.
This choice is because MergeGen is the most state-of-the-art tool tailored for code merging tasks so far.
Alternative choices like ChatMerge~\cite{chatmerge}. MergeBERT~\cite{MergeBERT} and DeepMerge~\cite{deepmerge} are not available in public.
We chose WizardMerge~\cite{wmerge} as the resolution suggestion provider to compare with \sys.
Although there are other suggestions providing tools such as RPredictor~\cite{RPredictor} and MESTRE~\cite{MESTRE}, they are not adequate because they are capable of predicting whether developers should retain the left version, retain the right version, or resolve the conflict by manually introducing updated code. 
However, they cannot provide suggestions for resolving conflicts that involve code block dependencies. 
As a result, they are not suitable for direct comparison with \sys.

\textbf{Dataset and Metrics Selection}
For RQ1 and RQ2, we used ConGra~\cite{congra} as the target dataset to evaluate the quality of auto-generated resolutions of LLM because of its large scale and diverse conflicts.
As Joern's C++ analysis ability is still not mature, we excluded the C++-based projects of ConGra to avoid biased results.
Besides, we also exclude the data that is marked as "textual" conflicts because these conflict exists in comments, where program analysis will not be effective.
Finally, the dataset includes 8,284 merge commit pairs.
For RQ3, considering WizardMerge can only analyze code written in C and needs to manually compile the whole project, comparing \sys with WizardMerge on ConGra's dataset is not practical because we will have to compile the two versions of code for each pair of conflicts, which will lead to an extremely large time cost.
Thus, we further utilize the C-based projects (\ie Linux and PHP) of WizardMerge's dataset used in its evaluation to answer RQ2 and RQ3.
As the models may output differently for the same input, we will repeat the experiment 10 times and calculate the average results to avoid biases.
Following ConGra~\cite{congra}, we adopt \textit{Edit Distance (ED)}~\cite{editdistance}, \textit{Winnowing Similarity (WS)}~\cite{winnowing}, and \textit{Cosine Similarity (CS)}~\cite{cosinesim} to measure code similarity at the character, structural, and semantic levels, respectively.
We do not rely on unit tests to validate the functional correctness of generated resolutions for two reasons: (1) many merged code fragments in our dataset are not accompanied by corresponding unit tests, and (2) performing functional validation at this scale requires prohibitive computational resources and time.
To mitigate the limitations of the string-based metrics, we additionally conduct manual validation on randomly sampled conflict resolutions when answering RQ2.

\subsection{RQ1: Comparison between MtCPG and CPG}\label{sec:rq1}

To address RQ1, we compare MtCPG with the original CPG in terms of graph size, measured by the number of nodes and edges.
We report statistics after graph–text alignment, during which unchanged nodes and edges are removed.
Since each merge case consists of two versions (i.e., \textit{VA} and \textit{VB}), we collect statistics for both.
The results are summarized in \autoref{tab:cpg_statistics}.
Overall, MtCPG consistently produces larger graphs than the vanilla CPG, as evidenced by higher average node and edge counts.
For the \textit{VA} version, MtCPG contains an average of 10,060.96 nodes and 40,119.42 edges, representing a 3.37\% increase in nodes and a substantial 24.96\% increase in edges compared to the vanilla CPG.
Similarly, for the \textit{VB} version, MtCPG reaches an average of 10,072.85 nodes and 40,187.03 edges, corresponding to increases of 3.35\% in nodes and 24.88\% in edges.
These results indicate that MtCPG preserves significantly richer dependency information, particularly in terms of inter-procedural and cross-change relationships, than the original CPG.

\begin{table}[]
\caption{The average number of node and edge from vanilla CPG and MtCPG. We use $\Delta$ Node and $\Delta$ Edge to represent the average increasement ratio of MtCPG compared with vanilla CPG.}
\label{tab:cpg_statistics}
\resizebox{\linewidth}{!}{\tiny
\begin{tabular}{l|llll|llll}
\hline
\multicolumn{1}{c|}{\multirow{2}{*}{\textbf{CPG Types}}} & \multicolumn{4}{c|}{\textbf{VA}}                                                                                                                             & \multicolumn{4}{c}{\textbf{VB}}                                                                                                                             \\ \cline{2-9} 
\multicolumn{1}{c|}{}                                    & \multicolumn{1}{c}{\textbf{Avg. Node}} & \multicolumn{1}{c}{\textbf{Avg. Edge}} & \multicolumn{1}{c}{\textbf{$\Delta$ Node}} & \multicolumn{1}{c|}{\textbf{$\Delta$ Edge}} & \multicolumn{1}{c}{\textbf{Avg. Node}} & \multicolumn{1}{c}{\textbf{Avg. Edge}} & \multicolumn{1}{c}{\textbf{$\Delta$ Node}} & \multicolumn{1}{c}{\textbf{$\Delta$ Edge}} \\ \hline
Vanilla CPG                                              & 9732.72                                & 32105.84                               & -                                   & -                                    & 9746.14                                & 32181.14                               & -                                   & -                                   \\
MtCPG                                                    & 10060.96                               & 40119.42                               & ↑3.37\%                             & ↑24.96\%                             & 10072.85                               & 40187.03                               & ↑3.35\%                             & ↑24.88\%                            \\ \hline
\end{tabular}}
\end{table}

Extracting additional code dependencies inevitably introduces extra computational overhead.
We therefore also measure the time required to construct MtCPG and the vanilla CPG.
On average, vanilla CPG construction takes 13.59 seconds, whereas MtCPG construction requires 17.31 seconds, resulting in a 27.37\% increase in construction time.
In extreme cases involving large graphs, vanilla CPG construction takes 8 minutes and 14 seconds, while MtCPG construction takes 15 minutes and 3 seconds.
This overhead is primarily attributable to the significantly larger graph size.
For example, in one large case, the vanilla CPG of \textit{VA} contains 1,552,767 nodes and 5,784,383 edges, whereas MtCPG contains 1,630,023 nodes ($\uparrow$ 77,256) and 6,969,369 edges ($\uparrow$ 1,184,986).
Overall, although MtCPG incurs higher construction costs than the vanilla CPG, the overhead scales proportionally with the amount of additional dependency information captured.
Since graph construction is performed offline and only once per merge scenario, this overhead is acceptable in practical settings.

\begin{figure*}[th]
	\centerline{\includegraphics[width=0.5\linewidth]{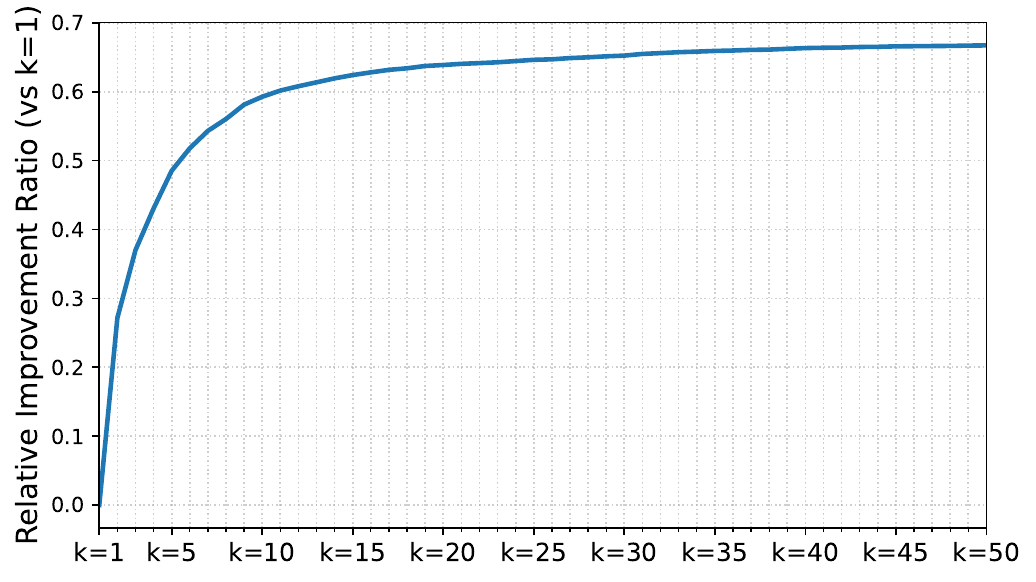}}
	\caption{\textmd{Relative improvement ration compared with \textit{k=1}.}}
	\label{fg:k-param}
\end{figure*}

We further evaluate the effectiveness of MtCPG for code context generation.
As described in \S \ref{sec:context_generate}, code contexts are generated using the MS-BFS algorithm over the graph.
Because traversing the entire graph is computationally expensive, we introduce a parameter \textit{k} to bound the exploration depth.
In this experiment, we vary \textit{k} from 1 to 50 and record the number of generated code contexts, as shown in \autoref{fg:k-param}.
For \textit{k} values greater than 1, we compute the relative improvement ratio by dividing by the increased number of contexts generated compared to when \textit{k=1}.
The results show that increasing \textit{k} yields substantial gains in context generation when \textit{k $\leq$ 10}.
Beyond this point, further increases result in diminishing returns with only marginal improvements.
Based on this trade-off between context coverage and computational cost, we recommend setting \textit{k $\leq$ 10} in practice.

\subsection{RQ2: Comparison among methodologies}\label{sec:rq2}

As \sys provides code dependency context to the LLM to enhance its ability to resolve conflicts.
To compare with it, we further treat the code adjacent to the specific conflict code block as context and set the context window length to 5, 20, and 50 lines of code, respectively.
These adjacent code contexts will be sent to the LLM and baseline model (\ie MergeGen) as suggestions to guide the conflict resolution generation.
According to the k-parameter selection results in \S \ref{sec:rq1}, we specifically tested values of k set to 1, 2, 4, 6, 8, and 10.
This allowed us to assess how varying thresholds impact the performance of \sys.
Additionally, we replace the MtCPG with CPG to conduct an ablation study to verify the effectiveness of MtCPG in providing helpful contextual code.
The evaluation workflow is as follows:
Firstly, the code contexts (either from adjacent code or generated by \sys) and the conflict will be fed to the LLM in the format of a prompt, and only the adjacent code context will be fed to the baseline model.
After obtaining the resolution output, the resolution will be compared with the ground truth in terms of accuracy by the three evaluation metrics.
%

% \usepackage{multirow}
\begin{table}[]
\caption{\textmd{Performance comparison on ConGra's dataset among \textit{MergeGen} (MG), vanilla LLM, \sys (RV) with CPG, and \sys with MtCPG. We use LoC to indicate adjacent contextual coode lines.}}
\label{tab:compare}
\resizebox{\linewidth}{!}{\tiny
\begin{tabular}{l|lll|lll|lll|lll}
\hline
\multicolumn{1}{c|}{\multirow{2}{*}{\textbf{System Names}}} & \multicolumn{3}{c|}{\textbf{C}}                                                                      & \multicolumn{3}{c|}{\textbf{Java}}                                                                   & \multicolumn{3}{c|}{\textbf{Python}}                                                                 & \multicolumn{3}{c}{\textbf{Overall}}                                                                \\ \cline{2-13} 
\multicolumn{1}{c|}{}                                     & \multicolumn{1}{c}{\textbf{ED}} & \multicolumn{1}{c}{\textbf{WS}} & \multicolumn{1}{c|}{\textbf{CS}} & \multicolumn{1}{c}{\textbf{ED}} & \multicolumn{1}{c}{\textbf{WS}} & \multicolumn{1}{c|}{\textbf{CS}} & \multicolumn{1}{c}{\textbf{ED}} & \multicolumn{1}{c}{\textbf{WS}} & \multicolumn{1}{c|}{\textbf{CS}} & \multicolumn{1}{c}{\textbf{ED}} & \multicolumn{1}{c}{\textbf{WS}} & \multicolumn{1}{c}{\textbf{CS}} \\ \hline
MG LoC=5                                                  & 54.92                           & 61.60                           & 79.04                            & 59.84                           & 69.79                           & 82.22                            & 56.59                           & 66.58                           & 80.50                            & 57.86                           & 65.77                           & 80.67                           \\
MG LoC=20                                                 & 53.63                           & 60.81                           & 78.99                            & 58.56                           & 69.29                           & 81.78                            & 52.10                           & 59.62                           & 80.53                            & 54.61                           & 63.47                           & 80.24                           \\
MG LoC=50                                                 & 52.35                           & 60.21                           & 78.91                            & 52.27                           & 62.70                           & 80.41                            & 48.95                           & 54.81                           & 78.23                            & 51.49                           & 59.88                           & 79.31                           \\ \hline
LoC=5                                                     & 58.51                           & 65.54                           & 80.64                            & 57.49                           & 68.71                           & 79.62                            & 62.92                           & 69.39                           & 80.05                            & 59.58                           & 67.65                           & 80.12                           \\
LoC=20                                                    & 61.82                           & 68.07                           & 80.99                            & 56.90                           & 68.01                           & 79.19                            & 63.94                           & 72.14                           & 80.69                            & 60.12                           & 69.01                           & 80.17                           \\
LoC=50                                                    & 59.48                           & 66.85                           & 80.67                            & 54.59                           & 66.82                           & 78.42                            & 63.33                           & 70.64                           & 79.30                            & 59.97                           & 68.73                           & 79.57                           \\ \hline
RV MTCPG k=1                                              & 66.42                           & 74.69                           & 86.95                            & 61.77                           & 72.50                           & 86.45                            & 72.59                           & 78.17                           & 89.37                            & 67.12                           & 75.16                           & 88.06                           \\
RV MTCPG k=2                                              & 66.60                           & 75.97                           & 87.17                            & 61.74                           & 72.72                           & 87.00                            & 73.39                           & 79.19                           & 89.32                            & 67.45                           & 76.10                           & 87.61                           \\
RV MTCPG k=4                                              & 67.46                           & 77.29                           & 87.03                            & 62.95                           & 73.89                           & 86.80                            & 73.75                           & 79.38                           & 89.55                            & 68.11                           & 76.23                           & 88.01                           \\
RV MTCPG k=6                                              & 65.29                           & 74.88                           & 85.99                            & 62.29                           & 72.51                           & 87.03                            & 73.90                           & 79.00                           & 89.68                            & 67.53                           & 75.59                           & 87.99                           \\
RV MTCPG k=8                                              & 66.62                           & 75.18                           & 87.08                            & 60.83                           & 72.24                           & 86.22                            & 73.27                           & 78.28                           & 89.57                            & 66.14                           & 74.66                           & 87.57                           \\
RV MTCPG k=10                                             & 64.61                           & 74.27                           & 85.66                            & 60.66                           & 72.10                           & 86.42                            & 70.84                           & 76.48                           & 88.14                            & 65.79                           & 74.40                           & 87.22                           \\ \hline
RV CPG k=1                                                & 64.61                           & 72.43                           & 84.69                            & 58.68                           & 68.37                           & 84.13                            & 68.35                           & 75.97                           & 85.94                            & 63.88                           & 72.77                           & 85.01                           \\
RV CPG k=2                                                & 64.19                           & 74.60                           & 84.05                            & 58.23                           & 68.97                           & 84.10                            & 70.28                           & 76.84                           & 86.45                            & 64.31                           & 72.83                           & 85.24                           \\
RV CPG k=4                                                & 64.04                           & 71.60                           & 84.81                            & 60.38                           & 70.87                           & 84.86                            & 70.88                           & 76.93                           & 86.99                            & 64.62                           & 73.36                           & 85.28                           \\
RV CPG k=6                                                & 63.41                           & 72.26                           & 84.25                            & 58.24                           & 68.88                           & 84.04                            & 70.99                           & 77.02                           & 87.02                            & 64.17                           & 72.92                           & 85.36                           \\
RV CPG k=8                                                & 61.39                           & 70.46                           & 83.94                            & 58.46                           & 69.35                           & 84.00                            & 71.07                           & 77.63                           & 86.47                            & 64.72                           & 73.39                           & 85.18                           \\
RV CPG k=10                                               & 62.88                           & 71.62                           & 84.09                            & 60.68                           & 71.28                           & 84.24                            & 71.90                           & 76.56                           & 87.64                            & 65.01                           & 74.38                           & 85.18                           \\ \hline
\end{tabular}}
\end{table}

The results in ~\autoref{tab:compare} show that \sys equipped with MtCPG achieves the highest accuracy in conflict resolution.
Among all evaluated configurations, \sys with $k=4$ (RV MtCPG $k=4$) delivers the best overall performance, achieving an ED score of 68.11, a WS of 76.23, and a CS of 88.01.
In comparison, the strongest standalone LLM baseline, which uses 20 adjacent lines of code as context (LoC=20), attains an ED of 60.12, a WS of 69.01, and a CS of 80.17.
This substantial performance gap demonstrates that the conflict-centric context retrieved by \sys is significantly more effective than naïvely supplying adjacent code lines.
MergeGen, which serves as another baseline, generally underperforms compared to the vanilla LLM setup, further highlighting the advantage of targeted context selection.
We also observe that providing more context does not necessarily lead to better conflict resolution quality, a phenomenon that holds for both adjacent-line contexts and graph-based code contexts.
Specifically, the performance of both the vanilla LLM and \sys initially improves as more context is provided (\eg, LoC=5 $\rightarrow$ LoC=20 and $k=1 \rightarrow k=4$), but subsequently degrades when the context size continues to grow (e.g., LoC=20 $\rightarrow$ LoC=50 and $k=4 \rightarrow k=10$).
We identify two primary reasons for this performance degradation.
First, increasing the value of $k$ expands the graph traversal scope, resulting in longer and denser code contexts.
Second, larger $k$ values introduce code dependencies that are farther away from the conflict location; such distant contexts are often weakly related to the conflict and therefore less effective in guiding resolution generation.
Together, these factors distract the model’s attention, ultimately harming resolution quality.
We illustrate this effect with a concrete example from the \texttt{cpython} project.
When setting $k=4$, \sys retrieves 92 code contexts comprising 2,477 lines of code.
However, increasing $k$ to 10 expands the context to 156 code blocks and 2,783 lines of code.
Crucially, most of the additional contexts are irrelevant to how to the conflict, as they only involve call-site invocations of the conflicting method without any data-flow dependencies (\ie, the method is invoked without arguments).
Consuming such longer but uninformative contexts misleads the model and degrades performance (ED: 65.56 $\rightarrow$ 57.22, WS: 71.48 $\rightarrow$ 64.90, CS: 79.04 $\rightarrow$ 70.30).

To assess the contribution of MtCPG, we further compare \sys using MtCPG with a variant that relies on the standard CPG (RV CPG).
As shown in \autoref{tab:compare}, the MtCPG-based configuration consistently outperforms the standard CPG across all $k$ values.
For example, at $k=4$, MtCPG achieves higher Edit Distance (68.11 vs. 64.62) and Winnowing Score (76.23 vs. 73.36).
These results confirm that MtCPG enables more effective context retrieval, which in turn leads to more accurate conflict resolution.
Interestingly, we observe that \sys's performance with vanilla CPG does not degrade as $k$ increases; instead, it largely plateaus and even shows a slight upward trend on the Python subset.
It can be attributed to the more conservative context extracted by CPG, which is typically confined to the same method and remains relatively short.
Such tightly scoped contexts preserve stronger relevance to the conflict and are less likely to distract the model’s attention.

A generated conflict resolution can be accepted only if 1) it is syntactically correct, and 2) its functionality is aligned with the ground truth resolution.
However, our current evaluation metrics are string-based and fall short in representing how far models can go in the above two aspects.
To mitigate the drawbacks it may cause, we did a sample validation to ensure the metric value and the resolution acceptance ratio are positively correlated.
In the validation, we first randomly sampled 300 conflicts, where 100 conflicts are \textbf{low} quality (\ie ED/WS/CS $<$ 33\%), 100 conflicts are \textbf{medium} quality (\ie 33\% $\le$ ED/WS/CS < 66\%), and 100 conflicts are \textbf{high} quality (\ie ED/WS/CS $\ge$ 66\%).
For each conflict and the code file it belongs to, we replaced the ground truth resolution code segmentation with the model-generated resolution in the code file to get the patched file.
Then, we verify whether the patched file can be accepted in terms of syntax-correction and semantic-correction.
To verify syntax-correction, we implemented an automatic syntax checker based on \texttt{-fsyntax-only} option of \texttt{clang}~\footnote{\url{https://clang.llvm.org/docs/ClangCommandLineReference.html}} for C, \texttt{Javalang}~\footnote{\url{https://github.com/c2nes/javalang.git}} for Java, and built-in AST parser for Python.
After a conflict's corresponding patched file passes the syntax check, we will manually check whether the patched file is semantically aligned with the ground truth file.
If the patched file passes the semantic check, we will mark it as an acceptable resolution.
We selected the samples based on results from \sys with \textit{k=4}.
As each conflict is resolved by the model 10 times, we only refer to the resolution with the highest overall metric.
We choose the resolution from \sys with \textit{k=4} and vanilla LLM with \textit{Loc=20} in this validation as they generally outperform other groups within the same approach.
Note that the metric-based random selection leans on the results of \textit{RV MtCPG k=4}, thus, only this group can guarantee that the selected metrics are evenly distributed.
For \textit{LoC=20} group, the numbers of high, medium, and low conflict resolutions are 84, 102, and 99, respectively. 
\autoref{tab:sytx_semt_check} shows the validation results.
It indicates that all metrics classification shows a high syntax pass ratio of resolution.
The failed cases are because the model accidentally outputs the code that should be placed outside of the conflict code blocks.
For example, if an invocation to a function includes 2 lines, while only the first line of it is covered by a conflict zone, the model should only output the resolution for the first line.
However, the model tends to fulfil the whole statement, thus outputting the entire invocation statement.
Therefore, the additional line will cause a syntax error and fail in the validation.
In the semantic validation stage, we found that high classification shows a stable pass ratio, while medium classification pass ratio will decrease by around 40\%-50\%.
For low classification, the pass ratio drops sharply to 2\%-3\%.
This experiment indicates that our evaluation metrics and the resolution acceptance ratio are highly positively correlated.

% \usepackage{multirow}
\begin{table}[]
\caption{\textmd{Syntax and semantic validation results. We use \textit{Sytx Pass} and \textit{Semt Pass} to refer to the number of resolution that passes syntax check and semantic check, respectively. We use \textit{Acc Ratio} to indicates the acceptance ratio of each metric classification.}}
\label{tab:sytx_semt_check}
\resizebox{0.8\linewidth}{!}{\tiny
\begin{tabular}{l|lll|lll}
\hline
\multicolumn{1}{c|}{\multirow{2}{*}{\textbf{\begin{tabular}[c]{@{}c@{}}Metrics\\ Classification\end{tabular}}}} & \multicolumn{3}{c|}{\textbf{RV MtCPG k=4}}                                                                                & \multicolumn{3}{c}{\textbf{LoC=20}}                                                                                      \\ \cline{2-7} 
\multicolumn{1}{c|}{}                                                                                           & \multicolumn{1}{c}{\textbf{Sytx Pass}} & \multicolumn{1}{c}{\textbf{Semt Pass}} & \multicolumn{1}{c|}{\textbf{Acc Ratio}} & \multicolumn{1}{c}{\textbf{Sytx Pass}} & \multicolumn{1}{c}{\textbf{Semt Pass}} & \multicolumn{1}{c}{\textbf{Acc Ratio}} \\ \hline
High                                                                                                            & 85/100                                 & 78/100                                 & 78.00\%                                 & 70/84                                  & 60/84                                  & 71.43\%                                \\
Medium                                                                                                          & 95/100                                 & 59/100                                 & 59.00\%                                 & 86/102                                 & 48/102                                 & 47.06\%                                \\
Low                                                                                                             & 87/100                                 & 2/100                                  & 2.00\%                                  & 80/99                                 & 3/99                                  & 3.03\%                                 \\ \hline
\end{tabular}}
\end{table}

\begin{figure*}[th!]
	\centerline{\includegraphics[width=\linewidth]{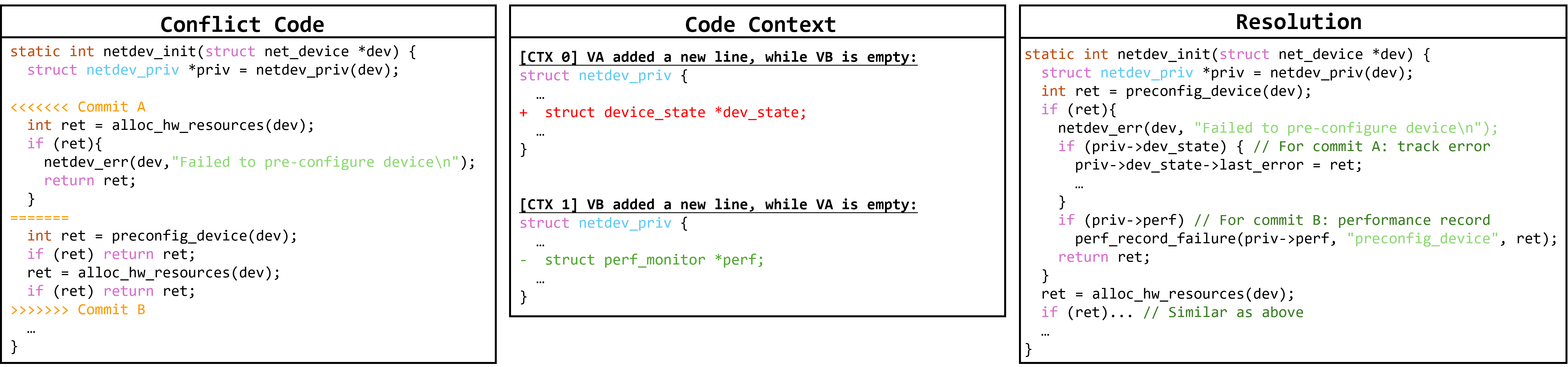}}
	\caption{\textmd{Case study of the effectivenss of \sys.}}
	\label{fg:case_study}
\end{figure*}

We demonstrate the effectiveness of \sys through a concrete case study in~\autoref{fg:case_study}.
Both \textit{Commit A (CA)} and \textit{Commit B (CB)} modify the function \textit{netdev\_init}.
Specifically, CA enhances error handling, whereas CB introduces a pre-configuration step prior to resource allocation.
Because these modifications overlap in the same code region, merging CA and CB leads to a conflict.
Lacking sufficient contextual understanding, an LLM fails to accurately capture the intent and scope of the two changes.
Since both edits appear individually reasonable, the LLM often resolves the conflict by selecting the code from either CA or CB, thereby discarding the other change.
In contrast, \sys analyzes code change dependencies across all modified code.
It identifies that, beyond the edits in \textit{netdev\_init}, CA also updates the structure \textit{netdev\_priv} by introducing a new field, \textit{dev\_state}, to track error states, while CB extends the same structure with an additional \textit{perf} field for performance monitoring.
Although these structural changes occur at different files from where the conflicts are located, and are not treated as direct conflicts during merging, they provide crucial contextual signals.
Leveraging this broader context, \sys infers that CA and CB introduce distinct and complementary functionalities that should be preserved as independent atomic changes.
By synthesizing these contexts, \sys successfully integrates the functionalities of both commits into a coherent resolution.
The generated result closely matches the ground-truth resolution (ED = 74.42, WS = 82.61, CS = 92.05) and is accepted as a valid conflict resolution.

% \usepackage{multirow}
\begin{table}[]
\caption{Comparison between LoC=20 and \sys with \textit{gpt-oss-20b} and \textit{llama3.1-8B-Instruct} models.}
\label{tab:different_models}
\resizebox{\linewidth}{!}{\tiny
\begin{tabular}{l|lll|lll|lll|lll}
\hline
\multicolumn{1}{c|}{\multirow{2}{*}{\textbf{System Names}}} & \multicolumn{3}{c|}{\textbf{C}}                                                                      & \multicolumn{3}{c|}{\textbf{Java}}                                                                   & \multicolumn{3}{c|}{\textbf{Python}}                                                                 & \multicolumn{3}{c}{\textbf{Overall}}                                                                \\ \cline{2-13} 
\multicolumn{1}{c|}{}                                       & \multicolumn{1}{c}{\textbf{ED}} & \multicolumn{1}{c}{\textbf{WS}} & \multicolumn{1}{c|}{\textbf{CS}} & \multicolumn{1}{c}{\textbf{ED}} & \multicolumn{1}{c}{\textbf{WS}} & \multicolumn{1}{c|}{\textbf{CS}} & \multicolumn{1}{c}{\textbf{ED}} & \multicolumn{1}{c}{\textbf{WS}} & \multicolumn{1}{c|}{\textbf{CS}} & \multicolumn{1}{c}{\textbf{ED}} & \multicolumn{1}{c}{\textbf{WS}} & \multicolumn{1}{c}{\textbf{CS}} \\ \hline
gpt-oss-20b LoC=20                                          & 80.73                           & 69.30                           & 85.24                            & 81.70                           & 70.05                           & 87.86                            & 84.23                           & 75.36                           & 88.68                            & 82.22                           & 71.80                           & 87.26                           \\
gpt-oss-20b RV                                              & 84.17                           & 74.23                           & 89.30                            & 85.29                           & 74.75                           & 91.39                            & 88.48                           & 78.89                           & 92.13                            & 86.05                           & 75.06                           & 91.61                           \\ \hline
Llama3.1-8B LoC=20                                          & 71.11                           & 65.27                           & 87.37                            & 74.08                           & 60.06                           & 85.90                            & 76.14                           & 69.67                           & 88.12                            & 73.78                           & 65.00                           & 87.13                           \\
Llama3.1-8B RV                                              & 65.96                           & 58.28                           & 81.46                            & 67.23                           & 50.33                           & 82.77                            & 67.56                           & 48.65                           & 77.83                            & 65.25                           & 49.08                           & 80.68                           \\ \hline
\end{tabular}}
\end{table}

To explore how \sys can impact other models, we evaluated \sys and LoC=20 with \textit{gpt-oss-20b} and \textit{Llama-3.1-8B-Instruct} respectively on the same dataset.
The results are shown in ~\autoref{tab:different_models}.
For \textit{gpt-oss-20b} model, \sys consistently outperforms the LoC=20 across all languages and metrics. 
For instance, the overall CS score improves from 87.26 to 91.61, and the overall WS score increases from 71.80 to 75.06. 
This indicates that \sys provides more relevant semantic information that helps the model resolve complex logical conflicts. 
Conversely, we observe a significant performance drop when applying \sys context to \textit{Llama3.1-8B-Instruct}. 
The overall WS score drops sharply from 65.00 to 49.08, and the ED score falls from 73.78 to 65.25. 
We manually checked 50 cases randomly selected from the low metric (\ie ED/WS/CS < 0.33) cases and found out that the key reason is \textit{Llama-3.1-8B-Instruct}'s misunderstanding of the code context.
Specifically, when the code context list is longer, the model will take a code context as the conflict to be resolved, and generate the resolution for that "conflict".
This suggests that lower parameter count (\ie 8B) potentially introduces a "reasoning bottleneck" where the smaller model fails to fully understand the conflict with long code context.

\subsection{RQ3: Comparison between \sys and other suggestion tool}\label{sec:rq3}

\subsubsection{Conflict Coverage}
We first assess how many conflicts out of the total conflicts can be handled by \sys and WizardMerge, respectively.
As tools designed to provide resolution suggestions, they inherently cannot guarantee the ability to analyze and generate relevant code suggestions for all types of conflicts due to design or implementation limitations.
If a conflict's context is not accessible, it implies that the LLM cannot leverage any assistance from the program analysis stage to resolve that particular conflict.
Therefore, we treat conflict coverage as an indicator of the generality and robustness of the approaches.
We use WizardMerge's dataset in this experiment.

\begin{table}[]
\caption{Comparison among LoC=20, WizardMerge's code context and \sys with k=4.}
\label{tab:other-tools}
\resizebox{0.8\linewidth}{!}{\tiny
% \begin{threeparttable}
\begin{tabular}{l|lll|l|l}
\hline
\multicolumn{1}{c|}{\multirow{2}{*}{\textbf{System Names}}} & \multicolumn{3}{c|}{\textbf{Overall}}                                                                & \multicolumn{1}{c|}{\multirow{2}{*}{\textbf{Conflict Cov.}}} & \multicolumn{1}{c}{\multirow{2}{*}{\textbf{Failure}}} \\ \cline{2-4}
\multicolumn{1}{c|}{}                                       & \multicolumn{1}{c}{\textbf{ED}} & \multicolumn{1}{c}{\textbf{WS}} & \multicolumn{1}{c|}{\textbf{CS}} & \multicolumn{1}{c|}{}                                        & \multicolumn{1}{c}{}                                  \\ \hline
LoC=20                                                      & 78.06                           & 79.56                           & 86.78                            & 343/343 (100\%)                                              &         N/A                                              \\
WizardMerge                                                 & 77.19                           & 70.65                           & 86.92                            & 311/343 (90.67\%)                                            &       \textcolor{red}{\ding{55}}  \textbf{$FR_1$}: 13; \textbf{$FR_2$}: 11;  \textbf{$FR_3$}: 8                                          \\
RV k=4                                                      & 80.14                           & 81.03                           & 90.46                            & 339/343 (98.83\%)                                            &   \textcolor{red}{\ding{55}}  \textbf{$FR_4$}: 4                                                    \\ \hline
\end{tabular}
% % \begin{tablenotes}
% % \footnotesize
% % \end{tablenotes}
% % \end{threeparttable}
}
\end{table}

~\autoref{tab:other-tools} presents the conflict coverage and explains why both \sys and WizardMerge encounter difficulties in handling certain conflicts. 
We used \textbf{$FR_{1-4}$} to represents: 1) Failure caused by not compiling the file that includes the conflict; 2) Failure caused by compiler's preprocessor removing or re-constructing code with the specific conflict; 3) Failure caused by the conflict being located within the comment; 4) Failure caused by the crash of static analysis.
Out of 343 total conflicts, WizardMerge successfully resolves 311 cases, achieving a coverage rate of 90.67\%, while \sys handles 339 conflicts, reaching an impressive 98.83\% coverage.
\sys fails to process 4 conflicts, all originating from the same conflict pair, due to crashes in Joern while performing initial program analysis on them. 
In contrast, WizardMerge fails in 32 cases, and these failures stem from several limitations.
First, WizardMerge can only analyze code that can be compiled into LLVM IR. 
If conflicts reside in code snippets that are not compiled (\eg, due to missing configuration settings for compilation options), WizardMerge cannot address them. 
Adding to this challenge is the fact that some compilation configuration options are mutually exclusive. 
For instance, when the Linux Kernel is set to target the x86 architecture, code for other architectures cannot be compiled, and vice versa. 
Consequently, such failures are unavoidable when the project includes these types of exclusive options.
Second, conflicts in conditional compilation code blocks (e.g., \textit{\#ifdef} and \textit{\#endif}) often go unprocessed by WizardMerge. This occurs because the C language preprocessor evaluates and removes code that does not satisfy the preprocessor conditions before generating the LLVM IR.
As a result, conflicts within these sections are discarded.
Finally, WizardMerge cannot manage code inside comment sections, as these are ignored during compilation, which is consistent with its reliance on compile-based analysis.
On the other hand, \sys operates at the source code level. 
By analyzing the raw code, \sys aligns more closely with the requirements of code merging, demonstrating greater robustness and universality compared to WizardMerge.

\subsubsection{Conflict Resolution Performance}

Similar to \S \ref{sec:rq1}, we conducted an experiment to evaluate the contextual suggestion effectiveness of WizardMerge and \sys respectively.
The evaluation workflow and the model selection are the same as in \S \ref{sec:rq1}, except that we only evaluated \sys with \textit{k=4}, LoC=20 and WizardMerge's contextual code as shown in ~\autoref{tab:other-tools}.
The findings demonstrate that \sys outperforms both LoC=20 and WizardMerge across all three metrics.
Specifically, \sys achieves the highest values for WS (81.03), ED (80.14), and CS (90.46), suggesting that it is the most effective method for conflict resolution in this dataset.
In comparison, the model with WizardMerge produces weaker results, particularly in WS (70.65), but its performance in CS (86.92) is competitive with LoC=20 (CS=86.78). 
Overall, \sys stands out as the most balanced and effective method, excelling across all key metrics for conflict resolution tasks.
In conclusion, we can state that \sys outperforms LLMs paired with WizardMerge in terms of auto-generated conflict resolution.

\section{Discussion}

% In this section, we discuss the limitations that exist in \sys’s design and explore the potential improvement room of
% \sys.

\textbf{Threats to Validity.}
In our evaluation, we utilized the ConGra dataset to demonstrate the efficiency of \sys across C, Java, and Python, which are among the most widely used programming languages.
However, due to the diverse syntax and unique features of different languages, this conclusion cannot be directly extended to other programming languages.
Additionally, we used three string-based metrics to evaluate the performance of the model's conflict resolution generation.
These three metrics cannot fully reflect the functionality alignment between the generated and ground truth resolution.
To mitigate this, we introduced a sampling validation to prove the positive correlation between the metrics and the resolution functionality alignment.

\textbf{Incomplete MtCPG Internal Definition.}
We have already presented the node and edge types of MtCPG in \S \ref{sec:mcpg}.
While these types are designed to cover the majority of scenarios in C, Java, and Python, they fall short when it comes to handling certain corner cases. 
For example, \sys currently cannot identify dependencies between try-catch-based exception handling and exception processing functions using MtCPG. 
Moreover, each programming language has its own unique features, and \sys is not yet capable of addressing dependencies arising from such language-specific characteristics. 
For instance, C supports inline assembly, which Joern treats as unprocessable code blocks. 
As a result, the control flow and data flow within inline assembly cannot be captured by MtCPG. 
In future work, we plan to further enhance MtCPG to better accommodate and process these diverse language-specific features

% \textbf{Inability to Handle Runtime-Dependent Program Dependencies.}
% %
% \sys's contextual suggestion generation is based on static analysis, which will be performed without actually executing the target program.
% %
% Therefore, static analysis results will compromise the accuracy due to the lack of runtime information~\cite{hybrid_analysis}.
% %
% For instance, in C programming, a function pointer is a pointer variable that holds the address of a function, enabling the function to be invoked through the pointer. 
% %
% Function pointers allow a program to dynamically determine which function to call at runtime. 
% %
% Since the target function of a function pointer may not be known at compile time, static analysis is unable to accurately identify the function the pointer will ultimately reference. 
% %
% As a result, the actual target of the function call cannot be determined during the compilation process. 
% %
% This introduces significant challenges for static analysis, particularly in complex control flows where function pointers can be assigned or modified at multiple points, making it impossible to predict the precise call path through static inspection.
% %
% In the future, we will consider combining static and dynamic analysis to handle dependencies from runtime.

\section{Related Works}

\textbf{Program Analysis Code Merging Approaches}
Unlike Git’s purely textual merge strategy, prior work has explored program analysis to enable semantics-aware code merging.
FSTMerge~\cite{fstmerge} represents an early semi-structured approach that combines structural merging for code elements with textual merging for string content.
Intellimerge~\cite{intellimerge} further applies semantic analysis by transforming source code into program element graphs and merging semantically matched elements.
SafeMerge~\cite{safemerge} employs verification techniques to analyze and combine independent edits, improving merge robustness.
WizardMerge~\cite{wmerge} assists developers by constructing dependency graphs for both conflicting and modified code blocks, highlighting code regions that influence or are affected by a conflict.
However, despite providing valuable insights, these program analysis–based tools still require manual intervention and do not achieve fully automated conflict resolution.
In contrast, \sys leverages program analysis to provide dependency-aware context to LLMs, enabling fully automated and more accurate conflict resolution.

\textbf{Machine Learning Code Merging Approaches}
Recent work has applied machine learning to automate conflict resolution.
MergeBERT~\cite{MergeBERT} formulates conflict resolution as a classification task over token-level merge patterns using a transformer-based encoder.
MergeGen~\cite{mergen} adopts a fine-grained, conflict-aware structural representation and generates resolutions using an encoder–decoder architecture.
ChatMerge~\cite{chatmerge} introduces a two-stage strategy that first predicts resolution strategies and then employs a large language model to generate conflict resolutions.
Unlike these approaches, which primarily rely on learned patterns or localized conflict representations, \sys integrates program analysis to systematically retrieve semantically relevant context, thereby achieving higher resolution accuracy.
\section{Conclusion}

We introduced \sys, a novel conflict resolution automation tool that combines the strengths of large LLMs and program analysis.
By constructing the MtCPG, \sys retrieves the contextual information relevant to a given conflicting code block.
It is subsequently synthesized with the conflict and used to guide the LLM in generating automatic conflict resolutions.
To evaluate \sys, we conducted experiments to assess its effectiveness in improving conflict resolution accuracy.
The results demonstrate that \sys outperforms standalone LLM by 13.29\% (WS), 10.46\% (ED), and 9.79\% (CS). 
Compared to the MergeGen baseline, \sys achieves improvements of 17.72\% (WS), 15.90\% (ED), and 9.10\% (CS). 
Additionally, \sys surpasses the combination of LLM with WizardMerge, achieving better performance by 3.82\% (WS), 1.85\% (ED), and 4.07\% (CS).
\section{Data Availability}

The source code related to the reproduction of \sys can be obtained from \url{https://figshare.com/s/fa83d33b5759da38a1a1}.

\clearpage
\bibliographystyle{ACM-Reference-Format}
\bibliography{citations}

\end{document}